# The Digital Architectures of Social Media: Comparing Political Campaigning on Facebook, Twitter, Instagram and Snapchat in the 2016 U.S. Elections

Michael Bossetta (University of Copenhagen)

## Abstract

The present study argues that political communication on social media is mediated by a platform's digital architecture – the technical protocols that enable, constrain, and shape user behavior in a virtual space. A framework for understanding digital architectures is introduced, and four platforms (Facebook, Twitter, Instagram, and Snapchat) are compared along the typology. Using the 2016 U.S. elections as a case, interviews with three Republican digital strategists are complimented with social media data to qualify the study's theoretical claim that a platform's network structure, functionality, algorithmic filtering, and datafication model affect political campaign strategy on social media.

The structural design of an environment – its architecture – intimately affects human behavior. This interplay between structure and agency is not limited to physical environs; it also applies to how users interact with, and within, online spaces. Scholars have argued previously that a digital platform's architecture can influence, for example: the norms of interaction among users (Papacharissi, 2009), the deliberative quality of their communication (Wright & Street, 2007), or their likelihood to enact democratic ideals (Freelon, 2015).

However, despite the rising interest in political campaigning on social media, few studies have questioned how a platform's design features influence political actors' communication strategies. This oversight is likely attributable to scholars' penchant for treating social media as a single media genre when in fact, these platforms exhibit significant differences in their network structures, functionalities, algorithms, and datafication models. The present study compares four social media platforms (Facebook, Twitter, Instagram, and Snapchat) along their digital architectures, with the aim of providing a new theoretical framework for studying political communication across social media platforms.

The scholarly inattention to the design features of social media is problematic for two reasons. First, political actors increasingly utilize social media as campaigning tools during elections. In the United States, political advertising on digital media across local, state, and national elections rose from 1.7% of ad spending in the 2012 election cycle to a 14.4% share in 2016 (Borrell, 2017). Moreover, a growing body of literature from countries outside the U.S. indicates that electoral campaigning on social media is a truly global phenomenon (Jacobs & Spierings, 2016; Strandberg, 2013; Grant, Moon, & Busby Grant, 2010). These and other case studies help elucidate how political actors use social media to advance their political agenda in a given social, cultural, or electoral context. Taken together, though, they lack a unifying theoretical framework for studying political communication on different social media platforms. This study provides such a model through its focus on digital architectures.

The second reason scholars' inattentiveness to the role of digital architectures is problematic concerns the increasing pluralization and fragmentation of the social media landscape. Newer platforms like Snapchat and Instagram vie for users' attention and encroach upon the market share previously held by platforms like Facebook and Twitter. In response, established providers either aggressively cannibalize the features of market challengers or, alternatively, attempt to buy them out entirely. Both Instagram and Facebook's incorporation of Snapchat-specific features, such as disappearing messages and self-documenting 'stories', exemplify the former strategy. The latter strategy, meanwhile, is evidenced by Facebook's acquisition of Instagram and WhatsApp, as well as Twitter's successful bid for Periscope (a live streaming service). The recent transformations in the social media landscape encourage political actors to adopt new platforms and features to reach different portions of the electorate. The existing trend among scholars to conduct single platform studies, or to subsume multiple platforms under a single "social media use" variable, is no longer sufficient to assess the complexity of contemporary "hybrid political communication systems" (Karlsen & Enjolras, 2016).

Aiming to assist future cross-platform research, this study is a theoretical piece offering a new heuristic for approaching political communication on social media. First, I propose a framework for conceptualizing digital architectures by presenting a typology that consists of four parts: network structure, functionality, algorithmic filtering, and datafication. The digital architectures of Facebook, Twitter, Instagram, and Snapchat (according to how they were structured in early 2016) are then compared along the typology. To bolster the comparison, two data types are incorporated in the study. The first is qualitative insights from interviews with three digital strategists working for Republican candidates in the 2016 U.S. election. The second is quantitative social media data from three platforms (Facebook, Instagram, and Snapchat). These empirical elements do not explicitly test the causal effect of digital architectures on campaign strategy; such an analysis is outside the scope of this paper. Rather, the empirical data is intended to help motivate new pathways for comparative cross-platform research that can, piece-by-piece, further our understanding of contemporary political campaigning.

**Digital Architectures and Affordances**

Whether an anonymous web forum like Reddit or 4Chan, a natively web-based social networking site like Facebook or Twitter, or an exclusively mobile app like Snapchat or WhatsApp, social media providers are faced with the challenge to develop digital communication tools that are easy to use and functional to the demands of varying user demographics. At the same time, these providers are competitors on the market and strive to develop different profiles that attract users, solicit advertisers, and sustain economic viability. Unsurprisingly, then, social media platforms display significant differences in their digital architectures: *the technical protocols that facilitate, constrain, and shape user behavior in a virtual space*. In line with what van Dijck and Poell (2013, pp. 5-6) refer to as the logic of "programmability," a social media's digital architecture is written in code, influenced by algorithms, and constantly tweaked by developers to maintain a competitive market advantage (see Lessig, 1999; Beer, 2009).

Previous scholarly work has argued effectively that digital communication technologies provide structural *affordances* to agents (Papacharissi & Yuan, 2011; boyd 2011). However, the concept of affordances is theoretically vague, and its analytical utility is questionable (Oliver, 2005; Parchoma, 2014). Broadly understood as "possibilities for action" (Evans, Pearce, Vitak, & Treem, 2017, p. 36), affordances lacks an agreed upon definition, and the highly inconsistent application of the term has been extensively critiqued elsewhere (Wright & Parchoma, 2011; Evans et al., 2017). As scholars work to refine the concept, there remains a need to "delineate how affordances work" (Davis & Chouinard, 2017, p. 6) by examining the underlying mechanisms of a technology and investigating how they shape user behavior. The argument here is that the architecture of a technology underpins its affordances, while offering a more empirically observable object of analysis.

Take, for example, stairs as a technology (Davis & Chouinard, 2017; McGenere & Ho, 2000). Stairs afford climbing, but it is the architectural design of stairs that influences their perceived and actual "climbability" (Warren, 1984). An affordance approach might consider the extent to which stairs enable climbing, whereas an architectural approach would examine how climbability is directly influenced by specific properties of the technology: the distance between steps, the angle of the rise, and other aspects relating to the structure's form. The two approaches are not necessarily at odds, but the architectural approach is arguably more conducive for comparing climbability across different types of stairs.

Applying the affordances concept to social media, Kreiss, Lawrence, & McGregor (2017, p. 12, original emphasis) have recently defined affordances as "what platforms are *actually capable of doing* and *perceptions of what they enable*, along with the *actual practices that emerge as people interact with them*." One could also argue that the capabilities, perceptions, and practices relating to a platform necessarily derive from its architecture. While the concept of affordances refers to what properties of communication are enabled by a platform (e.g., anonymity, persistence, or visibility [Evans et al., 2017, pp. 41-43]), the digital architectures heuristic drills into *how* a platform's specific design features affect particular communication practices. Put succinctly, digital architectures shape affordances and consequently, user behavior.

Apart from Kreiss et al.'s (2017) study, the application of the affordances concept to politicians' social media use is rare (see Stier, Bleier, Lietz, & Strohmaier, 2018 for a recent exception from Germany). This is most likely due to the fact that the large majority of studies on social media campaigning are single platform studies (Enli, 2017; Freelon, 2017; Filimonov, Rassman, & Svensson, 2016; Lev-On & Haleva-Amir, 2016; Kreiss, 2016; Jürgens & Jungherr, 2015; Graham, Broersma, Hazelhoff, & van't Haar, 2013; Vergeer & Hermans, 2013;

Larsson & Moe, 2012; Golbeck, Grimes, & Rogers, 2010). Most of the existing cross-platform analyses tend to cast their empirical gaze on citizens' discussion networks about political issues (Halpern, Valenzuela & Katz, 2017). This latter strand of research demonstrates that citizens' online communication about politics is influenced by how platforms are coded and designed. Halpern and Gibbs (2013), for example, show that the anonymity provided to user accounts on YouTube has a negative impact on the politeness of discussion in comment fields vis-à-vis the more personalized accounts required by Facebook. Dutceac Segesten and Bossetta (2017), meanwhile, find that in the social media discussions following the 2014 European Parliament elections, the Twitter publics of Sweden and Denmark were more closely aligned in their evaluations of Eurosceptic parties than users commenting on the Facebook pages of mainstream media outlets. They interpret their findings by arguing that similar user demographics are drawn to Twitter's specific features and news-oriented content profile (Perrin, 2015), creating a user base whose shared attitudes toward Euroscepticism override national variations between the two countries. Both of these studies suggest that the ingrained architectural features of a platform have direct implications for the types of political information and communication that flow across it.

Certainly, digital architectures alone cannot fully explain how or why political actors campaign on social media; the context of each race is critical in this regard (Auter and Fine, 2017; Aldrich, 2012). However, questioning how a platform's digital architecture influences campaign practices may provide insight into its strategy and, moreover, serves as a theoretical framework to inform comparative, cross-platform research designs. Additionally, the digital architectures heuristic is not limited to studies of political campaigning; it can also be applied to nearly any facet of online political communication: political debates among citizens, protest mobilizations, or journalistic reporting – to name a few.

In the following sections, four aspects of a social media's digital architecture are outlined: network structure, functionality, algorithmic filtering, and datafication. These categories have been chosen since each is argued to affect either the political content issued by politicians or citizens' access to political messages. Network structure influences how users identify and connect with political accounts. Functionality governs the rules of media production and diffusion across a platform. Algorithmic filtering determines what content users are exposed to, and datafication provides the means for politicians to target voters outside of their existing subscribers. These categories are not platform-specific and can therefore be used as bases for comparing politicians' digital strategy across different social media channels.

*Network structure*
The network structure of a social media platform refers to the in-built criteria governing connections between accounts. Almost by definition, "social" media allow individual users to connect and interact with peers: "Friends" on Facebook and Snapchat, "Followers" on Twitter and Instagram, or "Connections" on LinkedIn. Additionally, most social media allow users to establish connections with public figures, brands, or organizations (including political parties and politicians). Such high-resource actors typically maintain accounts with a different interface and suite of tools compared to the average user (e.g., Public Pages on Facebook or Business Profiles on Instagram).

Differences in the protocols underpinning network structure affect three aspects of user connections. The first is *searchability,* which refers to how users can identify new accounts and subscribe to their content (see boyd, 2011). The second is *connectivity*, referring here to how connections between accounts are initiated and established. Facebook's dyadic Friend structure, for example, requires peers to confirm relationships and has the effect of creating online networks that largely mirror a user's offline relationships (Ellison, Steinfield, & Lampe, 2007). Conversely, Twitter's connectivity is uni-directional by default and does not require a user to confirm a requested connection. This structural feature encourages one's Twitter network to be by-and-large composed of ties with no real-life connection (Huberman, Romero, & Wu, 2009).

The third aspect of network structure is *privacy*, which pertains to the ability of users to influence who can identify them through searches (searchability) as well as how connections interact (connectivity). Although Snapchat tends to encourage a more private network of close ties (Piwek & Joinson, 2016) compared to Instagram and Twitter's default open privacy settings, each platform allows users to customize whether incoming connection requests need to be approved by the user. Separately and together, the three elements of network structure – searchability, connectivity, and privacy – influence: the network topography formed on a platform, the strength of ties among users, and subsequently, the type of content likely to be generated on the platform (Bossetta, Dutceac Segesten, & Trenz, 2017).

*Functionality*
Functionality is the typology's broadest category and governs how content is mediated, accessed, and distributed across platforms. The first element of functionality is the *hardware* from which the platform is accessible: mobile, tablet, desktop, or wearable accessories like smartwatches and eyewear. Previous research suggests that hardware has direct effects on political content. Groshek and Cutino (2016), for example, find that differences in levels of civility and politeness in tweets correlate to whether they are issued from a desktop computer or mobile device. The second component of functionality is the layout of the *graphical user interface (GUI)*: the visual portal through which users access and interact with the platform's features. The GUI dictates the look of the social medium's home page, how a user navigates across different spaces within the platform (e.g., from a group page to an individual profile), and the available "social buttons" (Halupka, 2014, p. 162) that simplify processes of content diffusion across networks (e.g., Twitter Retweets or Facebook Shares).

Related to the GUI is the third category of functionality – the *broadcast feed*. The broadcast feed aggregates, ranks, and displays content on a platform in a centralized manner. Social media vary in terms of whether or not the platform maintains a centralized broadcast feed (such as the "News Feed" format popularized by Facebook), what types of accounts can contribute to the feed, and how content on the feed is accessed (i.e., scrolling down versus "click-to-open"). The fourth component of functionality is *supported media*. This refers to the multimedia formats the platform supports technically (e.g., text, images, video, GIFs), the size and length constraints placed on acceptable media (text character limits or video lengths), and the rules governing hyperlinking (both in terms of incorporating links from outside the platform as well as intra-platform linking via hashtags). Lastly, the fifth element of functionality is *cross-platform integration*: users' ability to share the same media across several platforms simultaneously.

These five components set the structural parameters for content creation and distribution across a network. Moreover, they are also mechanisms that give rise to user-generated norms of behavior influencing networks structures (i.e., how ties are maintained) and the content posted by users (what is customary and acceptable on the platform). A platform's functionality can "dispose networked publics toward particular behaviors" (Papacharissi & Easton, 2013, p. 176), and Vaterlaus, Barnett, Roche, & Young (2016, p. 599) have found that transgressing the "unwritten rules" of Snapchat can adversely impact interpersonal relationships among youths. To avoid similar negative effects with potential voters, political actors must be sensitive to the norms of appropriate content and interaction across different social media platforms. If they fail in their online performances though social media, political actors risk being perceived as out-of-touch, inauthentic, and subsequently a less electable to voters.

*Algorithmic Filtering*
Algorithmic filtering refers to how developers prioritize the selection, sequence, and visibility of posts (Bucher, 2012). For the typology's focus here on political campaigning, a distinction is made between *reach* and *override.* Reach describes how far a post cascades across a broadcast feed or set of networks, and algorithmic filtering can either promote or limit a post's reach. To drive revenue, many social providers allow users to *override* algorithmic filtering and further the reach of a post by offering pay-to-promote services, such as "boosting" on Facebook. Both reach and override are most relevant for social media platforms with one-to-many broadcast feeds (e.g., Facebook, Twitter, and Instagram). Other social media maintaining a predominantly one-to-one messaging profile – such as Snapchat, WhatsApp, Telegram, Kik, and Wickr – are less influenced by algorithmic filtering since messages are sent directly between users. When, though, the distribution and visibility of content is decided by algorithmic ranking, the coded operations implemented by developers have the power to shape users' shared perceptions of culture, news, and politics (Beer, 2009).

*Datafication*
Datafication, a term coined by Mayer-Schönberger and Cukier (2013), refers to the quantification of users' activities on a social media platform. Whenever users exercise the functionality of a platform, they leave digital traces (Jungherr, 2015) that can be collected for a variety of purposes: corporate advertising, market research, or internal refinement of a platform's algorithms by developers. According to the datafication logic, maintaining a social media profile during campaigns has less to do with establishing connectivity between politicians and citizens. Generally, levels of interactivity between these two actors on social media is low (Graham, Jackson, & Broersma, 2014; Jackson & Lilleker, 2011). The potential benefit for campaigns to take up social media electioneering is that they can monitor and harvest users' digital traces and appropriate them for decisions regarding persuasion or mobilization initiatives (Bimber, 2014). The 2012 Obama campaign, for example, effectively utilized data from Facebook through an application that encouraged supporters to send messages to friends who were calculated, based on multiple datapoints, to be persuadable (Kreiss & Welch, 2015).

The digital architectures typology distinguishes among three elements of datafication: *matching*, *targeting,* and *analytics.* Matching is the process of identifying users in a targetable audience through combining various forms of data. For political campaigns, digital strategists work in conjunction with polling firms to model audiences that are predicted to be favorable to a particular candidate or persuadable along a certain policy issue. Data from these models is then merged with party-collected data (i.e., voter files), data collected by the campaign, and third-party data purchased from commercial data warehouses that sell personally identifiable information (such as information from credit card companies). This data is used to build audiences of individuals who are first matched to their social media profiles and subsequently, targeted via the advertising services offered by the platform. Crucially for campaigns, analytics from these messages are interpreted in real time in order to "split-test" messages, and campaigns run thousands of randomized experiments to better craft and hone their message for persuasive effect.

## Data Collection and Method

With the four key features of the typology introduced, the digital architectures of Facebook, Twitter, Instagram, and Snapchat are systematically compared along each category in the following section. The comparison is informed by both qualitative and quantitative data. The former is primarily composed of interviews with three leading digital consultants from four Republican campaigns in the 2016 U.S. presidential election (Scott Walker, Rand Paul, Marco Rubio, and Donald Trump). Answering the call of Barnard and Kreiss (2013, p. 2057), interviews with campaign strategists were chosen to gain first-hand insight into how social media – and different platforms in particular – were utilized in relation to the overall campaign apparatus.

The interview participants included in the study are: Chasen Campbell, Vice President of Client Strategy at the Harris Media, the firm heading Rand Paul's digital strategy; Eric Wilson, Digital Director for Marco Rubio's campaign; and Matthew Oczkowski, Chief Digital Officer for Scott Walker's campaign and Head of Product at Cambridge Analytica, the digital consulting firm that assisted Donald Trump's general election campaign. The semi-structured interviews were conducted as part of the Social Media and Politics Podcast and are openly accessible for download via any podcast app.

To help illustrate the statements of the digital consultants, social media data from three of the four platforms (Facebook, Instagram, and Snapchat) is selectively presented. Twitter data was not collected during the timeframe studied, and limitations in Twitter's API rendered attaining comparable datasets for each politician unfeasible retroactively. The data that is included was posted between February 22 – March 15, 2016, a timeframe comprising one week before and two weeks after the string of primary elections known as Super Tuesday. This period has been chosen to ensure a high level of campaign activity on social media. The data stems from five campaigns' social media profiles: the three highest polling Republican candidates (Donald Trump, Ted Cruz, and Marco Rubio) and top two Democrats (Hillary Clinton and Bernie Sanders).

Facebook data from the politicians' public pages was collected using the rFacebook package (Barberá, Piccirilli, Geisler, & van Atteveldt, 2017) for the programming software R. Instagram data, on the contrary, is difficult to collect computationally since a user must receive special permissions from Instagram to harvest public data. To meet this limitation, Instagram data was collected via accessing platform's web version through the author's personal account. Politicians' Snapchat "stories" – compilations of user-generated messages that are accessible for 24 hours – were collected by utilizing Android emulation and screen capturing software. First, BlueStacks App Player was installed onto a Macintosh computer, enabling the author to access Android apps from the computer. After downloading Snapchat, the politicians' accounts were identified and followed, with the exception of Donald Trump. As explained in the paragraphs below, newcomers to Snapchat were difficult to identify, and for this reason Trump's account is not included in the analysis. However, another study (Al Nashmi & Painter, 2017) finds that over the same time period, the Trump campaign rarely sent Snapchats.

## Facebook, Twitter, Instagram, and Snapchat Compared

*Network Structure*

For a platform to be characterized as a social medium, it must support interactions among users. As argued above, network structure - the criteria governing connections between accounts - is a key component of a social media's digital architecture. Table 1 outlines the network structure characteristics of the four platforms.

|  | Network Structure | | |
|---|---|---|---|
|  | **Searchability** | **Connectivity** | **Privacy** |
| **Facebook** | High | Personal: Dyadic<br>Public Page: Uni-directional | Personal: Closed<br>Public Page: Open |
| **Twitter** | High-Medium | Uni-directional by default<br>Dyadic (by changing privacy) | Open by default |
| **Instagram** | Medium | Uni-directional by default<br>Dyadic (by changing privacy) | Open by default |
| **Snapchat** | Low | Dyadic by default<br>Uni-directional (by changing privacy) | Closed by default |

Table 1: Network Structure

A precondition for user interaction and network formation is *searchability* – how accounts are identified and their content accessed. On each of the platforms included here, political actors maintain publicly searchable profiles with openly accessible content. However, the searchability of political accounts varies across platforms and is be influenced by the account's username and elements of the graphical user interface. On Facebook, Twitter, and Instagram, the public pages of politicians are typically identifiable by simply searching their real names, and the authenticity of a page is often denoted via a blue verification checkmark on the GUI next to the account's username. For Instagram and to a lesser extent Twitter, searchability can be limited since multiple results (including parody accounts) are returned after searching a politician's name, and political accounts share the same format as that of the average user. On Facebook, politicians can establish public pages that set them apart visually (and functionally) from private accounts, and these pages feature prominently in search results. Political accounts on Snapchat have the lowest searchability and were extremely difficult to identify through direct search in the 2016 primaries. To follow a politician, users needed to know the exact username of a politician's account, which did not follow a uniform pattern (e.g., "GovernorOMalley", "CarlyforAmerica", and "Christie.2016"). The platform did not roll out a verification feature until November 2015, and most politicians did not have a verified account during the time under study.

In order to publicize their Snapchat accounts, campaigns focused on cross-platform promotion to their existing followers on other platforms. Wilson stated that the Rubio campaign promoted merchandise giveaways on Facebook and Twitter, where the campaign already had a strong presence. To be eligible, users were required to document that they followed Rubio on Snapchat by uploading screenshots from the platform to their other social networks. Oczkowski mentioned that Scott Walker, who had built a sizeable social media following through his Wisconsin recall election in 2012, promoted his Snapchat account across Facebook, Twitter, and Instagram but also would "plug it at events and rallies in person." While campaigns tried to popularize their lesser-known social accounts on other online platforms and at offline events, these messages would be primarily visible to the campaign's already existing supporters.

The other aspects of network structure - connectivity and privacy - are less relevant for political campaigning than they would be for analyses of individual user networks. On social media, citizens establish connections with political accounts in a uni-directional manner (that is, users subscribe to politicians' content without needing approval), since the privacy settings for these accounts are generally calibrated to be openly accessible. Thus, the campaigns did not exhibit significant differences in practices of connectivity or privacy across platforms.

As argued above, however, connectivity and privacy can affect the norms of communication among individual users. We can therefore expect that campaigns would be cognizant of these norms when crafting their communication strategy across different platforms. The low searchability, dyadic connectivity, and restrictive default privacy settings of Snapchat set it apart from more open platforms like Facebook, Twitter, or Instagram.

Likely, these features affect why Snapchat encourages a more informal mode of communication among close ties (Bayer et al., 2016). Oczkowski seems to confirm the informality and uniqueness of Snapchat communication when he states that the Walker campaign used the platform to "just give [followers] news and updates from the road on what we were doing, and making sure that we were using Snapchat appropriately and not just using it with the same exact content from every other channel."

Despite the different type of communication exhibited on Snapchat, the barriers to searchability limited the platform's utility for campaigns. Audiences were small, with Oczkowski estimating the Walker campaign's Snapchat following to be upwards of 10,000 and Wilson claiming the Rubio channel to get view rates of a "few thousand per day." In contrast, politicians on Twitter, Instagram, and particularly Facebook have a much larger user base, incentivizing campaigns to actively use the platform to reach voters. Comparing the view counts of the same videos posted across the platforms can give an indicator of the audience sizes that campaigns reach. A 30 second video posted by the Rubio campaign on March 5th, showing Rubio greeting supporters before a speech ahead of the Kansas caucuses, yielded 30,000 views on Instagram, 43,000 on Twitter, and 66,000 on Facebook – all significantly higher than the Wilson's estimation of the viewership on Snapchat. The number of Facebook video views registering highest is a consistent trend across the campaigns. For example, a video issued by the Trump campaign on March 13th – a 13 second video of Carly Fiorina denouncing Ted Cruz – garnered 676,000 views on Instagram, 778,000 on Twitter, and over 1.5 million on Facebook.

The massive user base of Facebook, whose platform allowed users to easily search and subscribe to politicians' accounts, renders the platform an attractive medium for campaigns to broadcast their message to a wide audience. At the time of the data collection, Facebook (2016) had 1.1 billion daily active users, Instagram (2016) approximately 300 million, and Snapchat around 120 million (Snap Inc., 2017). Twitter did not report daily active users at the time but claimed 310 monthly active users (Twitter, 2016). Although these are global figures and not limited to the U.S., Facebook clearly holds the pole position in regards to audience size (according to Campbell, 90% of American eligible voters). Figure 1 below depicts the number of posts issued on Facebook and Instagram, as well as the number of Snapchat stories.[1]

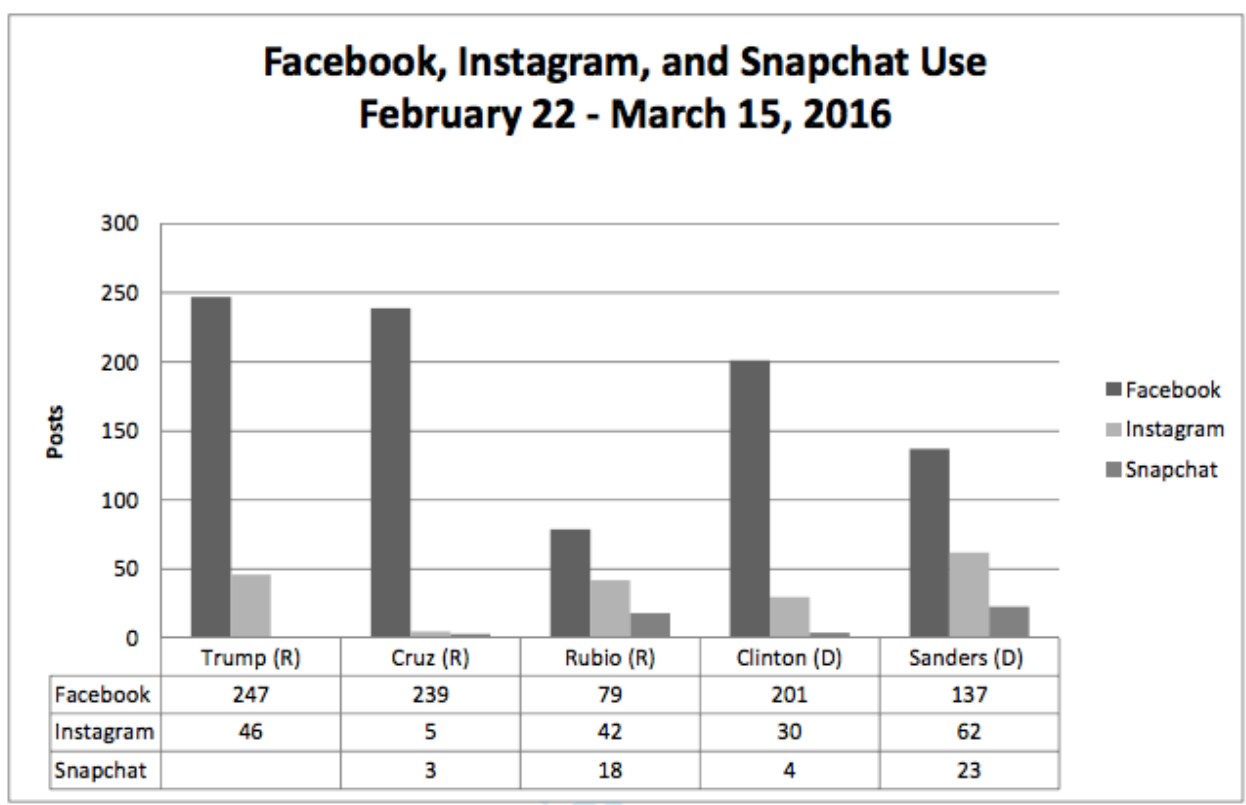

Figure 1: Facebook, Instagram, and Snapchat Use per Campaign

Unsurprisingly, of the three platforms included in Figure 1, campaigns posted the most content on Facebook. Figure 1 also shows that campaigns' propensity to use newer platforms like Instagram and Snapchat varied. Lower polling underdog candidates, like Rubio and

---

[1] The number of Snapchat stories, not individual snaps, are reported. Due to the ephemerality of snap messages, some may be missing from the collected data.

Sanders, showed high adoption rates for Instagram and Snapchat. However, the trend is not consistent as evidenced the Cruz campaign's low adoption rate.

*Functionality*

While network structure is one factor influencing Facebook adoption, the second part of the typology – functionality – also helps explain why campaigns take to Facebook. Table 2 outlines the differences in functionality across the three platforms:

|  | Functionality | | | | |
| --- | --- | --- | --- | --- | --- |
|  | **Hardware** | **GUI** | **Supported Media** | **Broadcast Feed** | **Cross-Platform Integration** |
| **Facebook** | Desktop, Smartphone, Tablet, Smartwatch | High Complexity, (E.g., News Feed, Public Pages, Groups, Events) | Text (63,206 characters) Images Video (45 minutes) Hyperlinks Hashtags | News Feed | None Supported |
| **Twitter** | Desktop, Smartphone, Tablet, Smartwatch | Medium Complexity (Can be broadened with dashboards) | Text (140 characters) Images Video (30 seconds) Hyperlinks Hashtags | Home Timeline and Highlights (opt-in) | None Supported |
| **Instagram** | Same as Facebook | Medium Complexity | Text (2,200 characters) Images Video (60 seconds) Hyperlinks (In Bio) Hashtags | Friend Feed and Explore Feed | Posting allowed to Facebook and Twitter |
| **Snapchat** | Smartphone Exclusively | Low Complexity, Simple Layout | Text: (under 31 charactures) Images Video (10 second limit) | Story Feed and Discover Feed | None Supported |

Table 2: Functionality

The first aspect of functionality is hardware. Facebook, Twitter, and Instagram are accessible from multiple types of hardware: desktop computers, tablets, smartphones, and smartwatches. Snapchat, on the other hand, is *exclusively* mobile and cannot be accessed from any other type of device. This hardware-specific feature of Snapchat has two direct implications for content creation on the platform. First, in order to post content featuring a political candidate, the person filming snaps from a smartphone must be in close physical proximity to the candidate. The digital directors stated that a candidate's "body man," or personal assistant who travels with the candidate, was usually responsible for the Snapchat account. The second implication of Snapchat's mobile exclusivity is that content needs to be uploaded directly from the mobile device, and therefore little editing or consultation with the campaign occurs before publishing content to a story. On the other platforms, by contrast, campaigns have the ability to upload edited content at scheduled, strategic time points. Wilson hints at how Snapchat's digital architecture generates a type of content different than on other platforms:

> "The unique thing about Snapchat is it *has* to be done right there. You can't upload a photo, you can't edit a video; it has to be physically from that device. So, you were seeing stuff that was coming right from, you know, where Marco was at that exact moment. It wasn't coming back to headquarters and getting filtered or edited in any way."

Since Facebook, Twitter, and particularly Instagram provide several functions to edit content prior to publishing, the type of visual content on these platforms is generally more polished and complex (i.e., infographics or memes). Figure 2 below illustrates how Snapchat's hardware restrictions encourage a more raw type of footage, versus Instagram's more artistic, edited shots. Both posts were published on February 26 and cover the same event. The left is a screenshot of a Snapchat video from a rally while the right depicts the campaign's Instagram representation of the event through a still image.

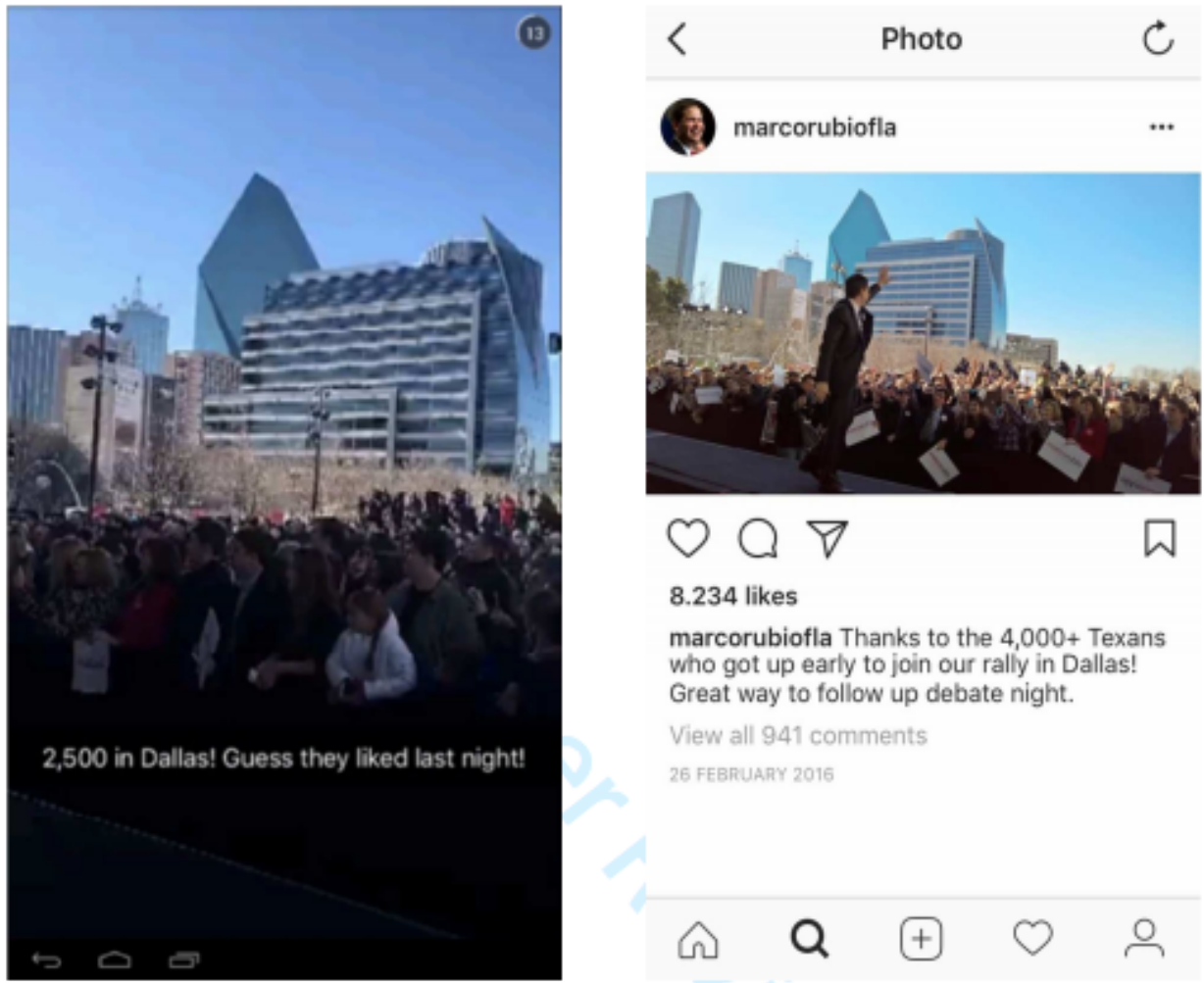

Figure 2: Snapchat/Instagram Comparison

Clearly, the Instagram photo has been edited (i.e., "filtered") for artistic effect. Moreover, the picture has been strategically chosen to show both the candidate and a band of enthusiastic supporters. On Snapchat, the audience is depicted in real time and appears much more mundane. Interestingly, the two representations also differ in the number of reported attendants at the rally (2,500 on Snapchat versus 4,000 on Instagram). This difference may signal that the ability to control or schedule content allows campaigns more time to validate or correct information.

Snapchat's less filtered glimpses into the campaign, compared to the other platforms' more polished visual content, is thus not only attributable to hardware but also its *supported media,* outlined above in Table 2. All four platforms supported text, images, and video, but they placed different constraints on the length of these media at the time of the campaign. Concerning text, Facebook capped posts at 63,206 characters, Twitter its notorious 140, Instagram limited captions to 2,200 characters, and Snapchat only allowed 31 characters to be overlaid to an image or video "snap." Regarding video, Facebook supported content up to 45 minutes, Twitter and Instagram a much lesser 30 and 60 seconds respectively, and Snapchat only 10 seconds per snap. Uploaded images are supported on Facebook, Twitter, and Instagram, although the optimal pixel size and level of compression varies across them. This means that if a campaign wants to share the same image across different platforms, creative teams may be enlisted to alter the image to meet the requirements ingrained in the platform's architecture.

The types of multimedia the platform supports, and the limitations placed on them, directly affects the content campaigns can communicate. Although Instagram and Twitter supported video, their limitations on length do not allow for substantial content from debates or media appearances. Video content on Instagram was scant, with videos comprising a proportionately low percentage of posts compared to images. The percentage of video content on Instagram, by campaign and in descending order, was: Trump (15%), Rubio (10%), Sanders (4%), Clinton (3%) and Cruz (0%). Facebook had a much higher percentage of video content, with most running over 60 seconds.

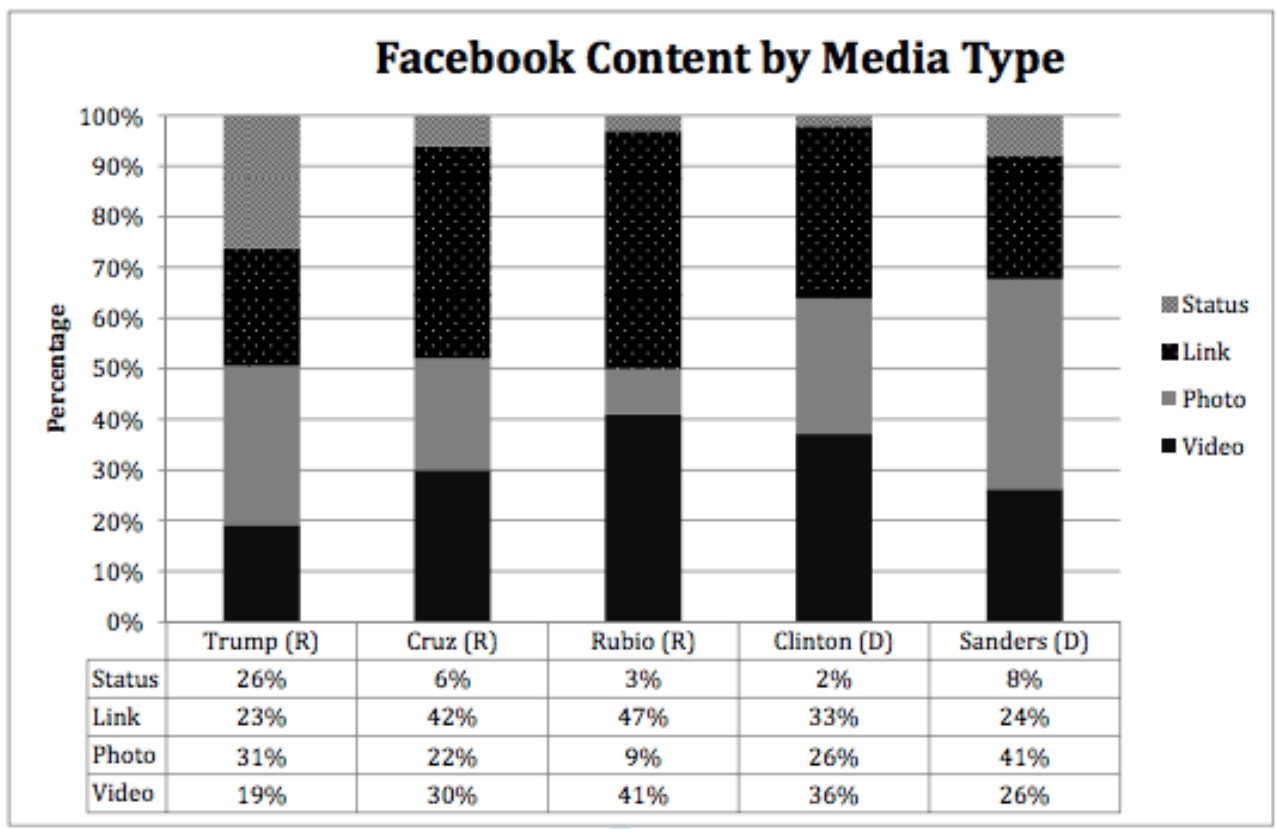

Figure 3: Facebook Content by Media Type

Supported media also refers to the rules governing hyperlinking, and Figure 3 shows that between 23%-47% of campaign's Facebook content comprised of links. By-and-large, links were aimed at redirecting users to the campaign's website or to a media article about the candidate. Although limitations in the data do not support a strict comparison, similar usage of links can also be expected on Twitter. On Instagram and Snapchat, campaigns could include web addresses to their posts in text, but they were not actionable (i.e., users could not click on them to be directed off the platform). One exception is that on Instagram, an actionable link can be included only in a user's profile description. This led the Clinton and Rubio campaigns to encourage users to "check out the link in bio for more info." The purpose of driving users off the platform and onto the candidate's site is to sign them up for email lists. Oczkowski described emails as "the lifeblood of fundraising" since "over 70% of all money raised online comes from email programs," and they're also "very helpful in turning people out to events and rallies."

How users access media content within these platforms, though, is influenced is by two aspects of functionality: the *broadcast feed* and the *graphical user interface.* Whereas the former structures content, the latter governs how it is displayed. Facebook's centralized broadcast feed (i.e., the "News Feed") provides the user with a series of algorithmically filtered content published by peers, subscribed pages, advertisers, and other sources appearing on the feed as a result of algorithmic contagion. Twitter's centralized feed ("Home timeline") presents users with chronologically-ordered posts based on their subscriptions. On mobile devices, users also can opt-in to the Highlights feed, which presents users with more algorithmically filtered content based on relevance. Instagram has two broadcast feeds: one for subscribed connections (and advertisers), and the "Explore" feature that provides content suggestions to users. Snapchat's digital architecture, by contrast, includes almost no algorithmic filtering; the platform sorts content chronologically according to when a connection posted a message. Snapchat does, however, have a mass broadcast feed in the form of "Live Stories": series of user-generated content that are curated by the platform and typically focused around an event or geographical location.

So far, the functionality of the platforms has been compared according to how elements of their digital architecture influence content production and diffusion *within* a platform. The last component of functionality relates to *cross-platform integration*: whether users can share the same content across different platforms simultaneously. Neither Facebook nor Twitter allows posting to different platforms, but Instagram allows users to share posts across Facebook and Twitter simultaneously. On Snapchat, users can only save content taken in the app's camera and repurpose it to other platforms. Since the same content can be shared across Facebook, Twitter, and Instagram, and content taken via Snapchat can be uploaded to these platforms as well, it cannot be assumed that political campaign's content is specific to any one particular platform. For example, both the Trump and Rubio campaign uploaded Snapchat videos (1 and 2, respectively) onto their Instagram accounts. Hillary Clinton uploaded a picture of one of her tweets to Instagram. The high percentage (26%) of text only statuses making up Donald Trump's Facebook content, as shown in Figure 3, were largely comprised of the same messages he posted on Twitter.

Thus, although a platform's architecture might encourage or necessitate a certain type of content, scholars should not assume that political content issued on a social media platform is necessarily specific to it. To illustrate this point empirically, Figure 4 presents the percentage of Instagram content that was also present on Facebook. The "Direct Overlap" category represents when the visual content *and* caption were the exact same across both platforms. "Edited Overlap" refers to when the visual content was the same but the caption was changed (for example, to incorporate a hashtag, change a hyperlink, or slightly modify phrasing). "Instagram Only" is the percentage of content that was not posted to Facebook.

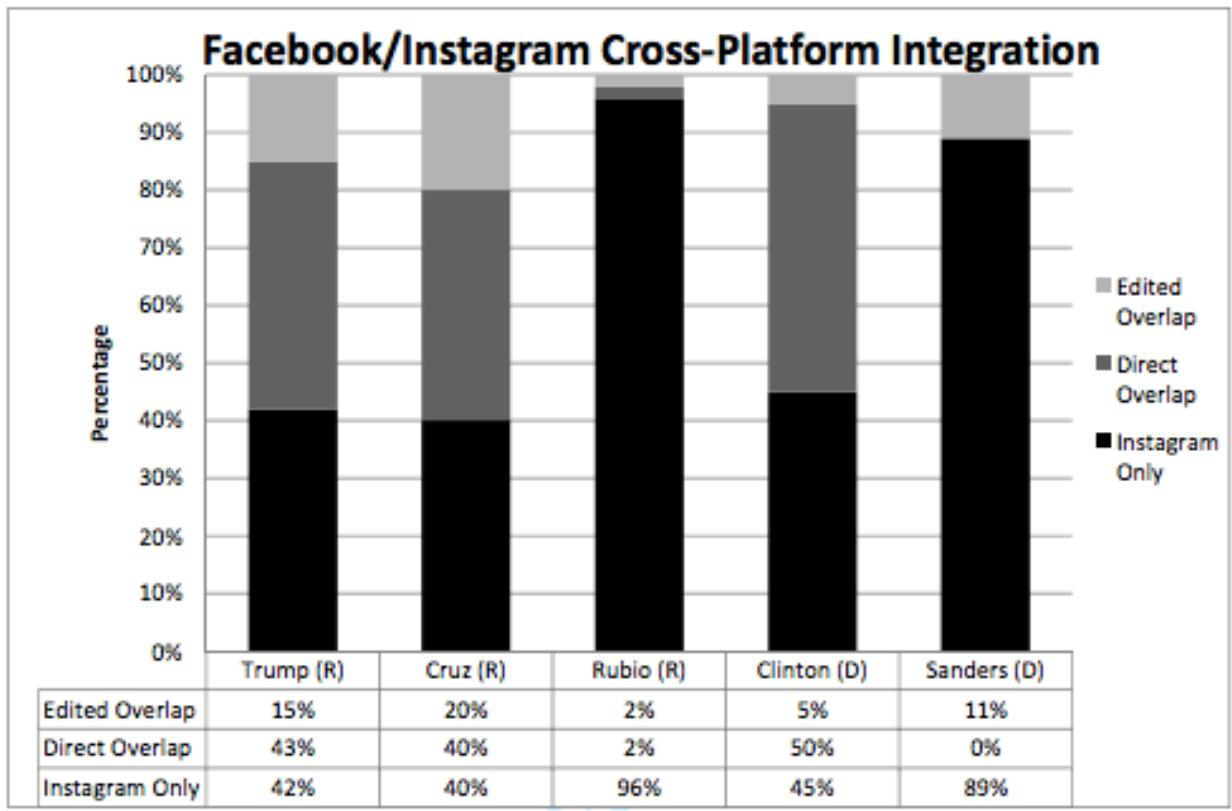

Figure 4: Facebook/Instagram Cross-Platform Integration

Figure 4 reveals that for three out of the five politicians (Trump, Cruz, and Clinton), over half of the content posted to their Instagram profiles was also made available on Facebook. For Rubio and Sanders, on the other hand, content posted on Instagram was typically not uploaded to Facebook. These two underdog campaigns were also the most active on Snapchat, suggesting that new platforms may be more attractive to low-polling campaigns.

*Algorithmic Filtering*

The remaining two categories of the digital architectures typology - *algorithmic filtering* and *datafication* – are difficult to assess with public social media data, but they are presented briefly here to round off the comparative platform analysis. Table 3 presents an overview of the similarities and differences across platforms.

|  | Algorithmic Filtering | |
|---|---|---|
|  | **Reach** | **Override** |
| **Facebook** | Heavily filtered (relevance) | Pay to promote<br>User-diffusion (Sharing) |
| **Twitter** | Moderately filtered (chronology) | Pay to promote<br>Index via hashtags<br>User-diffusion (Retweeting) |
| **Instagram** | Moderately filtered (chronology) | Pay to Promote<br>Index via hashtags |
| **Snapchat** | None | No algorithm to override |

Table 3: Algorthmic Filtering

As alluded to previously, Facebook's broadcast feed exhibits heavy algorithmic filtering based on calculated relevance, while Instagram and Twitters' algorithms place more emphasis on the chronological order of posts. Snapchat has little to no filtering, granting the user a high level of autonomy in selecting content.

Algorithmic filtering directly influences the organic (i.e. non-paid) *reach* of a post. Facebook page posts, for example, typically reach less than 10% of subscribers organically, a number that continues to decline over time (Manson, 2014). The algorithms of Twitter and Instagram, favoring chronology over relevance, grant campaigns a more direct line to subscribers. However, filtering by chronology also makes the reach of the post sensitive to the overall activity on the platform. During times of heightened political activity (e.g. around an election or debate), posts can be easily "drowned out" by higher levels of posting by other users. Snapchat's virtually non-existent filtering allows users the most direct access to campaign content, with the important caveat that these broadcasts disappear after 24 hours.

To counter these limitations and extend reach, each platform offers mechanisms to *override* algorithmic filtering. Facebook, Instagram, and Twitter offer pay-to-promote services to extend the reach of an existing post such "boosting" to a wider audience based on demographics or interests. Apart from this market-driven feature, campaigns can enlist the help of supporters to diffuse messages across their own networks on Facebook and Twitter (via sharing and retweeting). On Twitter and Instagram, hashtags are an effective means to index posts outside of one's immediate follower network (Facebook has also incorporated hashtag functionality, although it remains largely ineffective for increasing reach due to Facebook's less open network structure). Although Snapchat lacks a curating algorithm to be overridden, being featured in a Snapchat "Live Story" can drastically increase the reach of their content. Wilson mentioned that Snapchat worked with campaigns to promote candidacy announcements, debate days, and election days. When the Rubio campaign was feature in a Live Story, which were broadcast either nationally or in a specific state, view counts would go from the average "few thousand per day" to "definitely get[ing] up into the higher five figures of views." Whereas campaigns can utilize override mechanisms to *extend* the reach of a post, they generally rely on datafication techniques to *control* the audiences of specific posts.

*Datafication*

Datafication, in a campaign context, implies the process of quantifying users' activity for strategic purposes. On the one hand, data is utilized for *matching* and *targeting* specific audiences with the intent of persuasion or mobilization. On the other, datafication allows for campaigns to monitor and collect *analytics* that help inform future strategy. Datafication is a complex, expensive, and iterative process in contemporary digital campaigning. Oczkowski describes the process as, firstly, using a combination of data from voter files, commercial warehouses, and polling from a small part of the electorate (around 1,500 people) to then, secondly, extrapolating this data to build look-alike audiences of larger portions of the electorate. Targeted messages are then issued to persuade voters, and *analytics* (often monitored in real time) help measure their effectiveness. Oczkowski describes the process while hinting at the iterative character of datification:

"So, I say, these are Trump supporters, these are people who love to reduce taxes, these are gun supporters, these are the religious rights – all based on survey data and database data that I have and that I've brought in. From there, we're then segmenting audiences for the purposes of our media teams to buy digital ads or to buy television, but also for creative teams to be able to craft messages: the ads, the types of things we're saying to people. Those two things then come together, we spend money to do paid media, and then we go back in the field and we're consistently polling to see if what we're doing is working and how effective it is."

The above quote highlights how datafication has both offline (traditional polling and television) and online (digital databases and ads) dimensions. Regarding the present study's focus, the digital architectures of each platform offer varying types and degrees of datafication, which are summarized below in Table 4.

|  | Datafication | | |
|---|---|---|---|
|  | **Matching** | **Targeting** | **Analytics** |
| **Facebook** | Highly developed "Custom" and "Lookalike" Audiences | Extremely sophisticated Several Ad formats | Complex, Real time analytics (Walled-Garden) |
| **Twitter** | Moderately developed "Tailored Audiences" | Moderately sophisticated Few ad formats Tagging journalists | Open API Dashboards |
| **Instagram** | Same as Facebook | Same as Facebook | Same as Facebook |
| **Snapchat** | Least developed "Snap Audience Match" (opt-out) | Least sophisticated Ads in Stories (opt-out) | Rudimentary in primary, improved in general election |

Table 4: Datafication

Matching, or the process of linking data to online social media profiles, differs across platforms. Campbell describes the high sophistication of Facebook's matching service, "Custom Audiences," as being able to match 70-80% of users in a database within 30 minutes based solely on their names and home mailing addresses. Once a custom audience is built, Facebook can recommend other users who are outside of the custom audience, but calculated to share similar datapoints, through the "Lookalike Audience" feature. Matched or lookalike audiences can then be targeted via a plethora of ad formats customizable by: multimedia, placement on the GUI, and hardware (mobile versus desktop). Owned by Facebook, Instagram offers the same suite of tools. Twitter has a similar matching and lookalike service called "Tailored Audiences." However in comparison with Facebook, Twitter's matching is less sophisticated (e.g., it does not support home mailing addresses) and offers few ad formats outside of promoted tweets, accounts, and trends. According to Campbell, though, Twitter is used to target lists of known journalists so that: "the people who are writing the [mainstream media] stories at the end of the day are the ones seeing your ad, and you're encouraging earned media responses." Snapchat, as the newest platform with the least developed datafication features, only began offering audience matching ("Snap Audience Match") in September 2016, one month before the general election. Targeted ads on Snapchat are inserted between stories, and the platform offers users the option to opt-out of matching and targeting in their privacy settings.

Both matching and targeting are resource-intensive processes involving extensive knowledge and monetary capabilities. As highlighted by Kreiss & McGregor (2017), technology firms offer consulting services to high-profile campaigns to assist them in crafting their targeting strategy. Campbell highlights the importance of these services when he states:

> "We value those relationships and there are some very, very smart people working at these companies that are helping us to execute the strategy that we're coming up with, and in some cases even help us form the strategy that we're coming up with, because they understand their platforms better than anyone does…almost daily, we're speaking to our teams [at Google, Facebook, and Twitter] that actually help to facilitate all of the advertising".

While tech companies have partisan teams that assist campaigns in their targeting strategies, this relationship is ultimately symbiotic: companies raise revenue, campaigns raise electoral support. For campaign consultants, analytics become crucial for assessing the effectiveness of a communication strategy and necessary for acquiring more resources for digital advertising. As Wilson remarks, "It's hard to make the case for resources when you don't have the analytics to back it up." Analytics help measure return on investment (ROI), but the availability of analytics differs across platforms.

Facebook has increasingly taken steps to limit access to both Facebook and Instagram data; the platform's "walled-garden" approach requires payment (via advertising) in exchange for data. According to the interviewees, Snapchat as a start-up was largely unable to inform campaigns about their view rates, and the purpose of advertising on the platform was simply to better get a sense of engagement. Twitter, according to Oczkowski, "is really the only open Firehouse left," and Wilson mentioned using Twitter to monitor mentions of certain initiatives the Rubio campaign was running, such as a "Vote Early Day" initiative aimed to increase turnout. Dashboard applications like TweetDeck or Hootsuite can campaigns help monitor and measure specific initiatives. However, Oczkowski also stated the limitations of Twitter data: "Twitter data's great but it doesn't represent most voters in America; it's a minority of very vocal people." In order to understand and reach a larger portion of the electorate, campaigns must invest significant resources into both online and offline data acquisition. Moreover, it must be stated that from a data collection and targeting standpoint, social media platforms comprise only a part – but an increasingly important part – of the modern day campaign apparatus.

## Discussion and Conclusion

Although the social media landscape remains dominated by early market entrants like Facebook and Twitter, scholars need new approaches to meet – but also anticipate – rapid changes in this ever-evolving digital space. The present study has put forth the argument that scholarly attention to a platform's digital architecture provides a valuable and flexible heuristic to approach cross-platform research of social media. Ultimately, the study's aim has been to illuminate new pathways for comparative social media research in the context of political campaigning, but the framework can also be applied to studies of citizens' discussions or journalistic reporting.

Theoretically, the study posits that four aspects of a platform's digital architecture influence political communication on social media – network structure, functionality, algorithmic filtering, and datafication. Respectively, these four infrastructural elements of platform design impact the decisions that political campaigns make in terms of: the audiences they try to reach, the form and content of messages they produce, the diffusion patterns of these messages, and how financial resources are allocated for digital campaigning on social media.

The study's exploratory operationalization of the digital architectures framework, applied to the case of the 2016 U.S. elections, yields three interesting results. First and foremost, campaigns shared much of the same content – in text, images, and video – across different social media platforms. Basing their study on interviews with U.S. campaigners, Kreiss et al. (2017, p. 2) argue that "campaigns must produce their own creative content for very different platforms like Facebook, Instagram, Twitter, and Snapchat." While certainly true to an extent, this study – even with its limitations - finds an overlap in campaign messages across all of the platforms studied. Although one platform may encourage (or even necessitate) a certain type of content, other platforms with similar functionalities can support the re-appropriation of content across multiple channels. Scholars should therefore exert caution in assuming that the content posted to a particular social media is unique to the platform. Cross-platform analysis, with rigorous attention to platforms' digital architecture, can help ascertain whether and why content is specific to a given platform.

Second, both the interviews and social media data point to the dominance of Facebook in the 2016 election cycle. The platform was the most attractive social media for political campaigns on account of several architectural design features. Facebook's public pages, providing an open network structure with easily searchable accounts, supported large social media followerships (demonstrated here, for example, by differences in video view rates across platforms). The functionality of hyperlinking, meanwhile, was heavily utilized by campaigns to drive traffic to their websites (for fundraising) and collect emails (for audience matching). Non-restrictive rules regarding video lengths rendered the platform a key medium for long-form visual telecommunication. Algorithmic filtering, and the ability to override it via paid advertising, allowed campaigns to reach potential voters outside of their organic follower bases. Moreover, Facebook's sophisticated matching, targeting, and analytics suites enabled high-resource campaigns to split-test messages to voters in strategic geographical locations.

Third, even though campaigns invested less heavily in newer platforms like Instagram and Snapchat, the study finds that all candidates analyzed were active on these platforms. A standard trend observable across the campaigns is that Instagram was used more often of the two. This is likely due to the functionality differences between the two platforms: Instagram allows campaigns to control the image of their candidate via uploading polished content at a scheduled time. Snapchat, while carving its niche in the social media marketplace through it's live and disappearing broadcast features, was likely more risky (and less useful) for campaigns to adopt than Instagram. Crucially, Snapchat lacked a comprehensive datafication incentive to reward politicians' who invested in the platform. Future work can dive deeper into investigating the content (and timing) of messages on these and other emerging platforms, in order to investigate whether they reveal patterns of communication that help elucidate a campaign's wider strategy.

The empirical analysis is, certainly, limited by several factors. Twitter data was not attainable, and the data from other platforms is solely that which was publically available. Targeted advertisements are often unpublished, rendering their collection via traditional computational means difficult. Such private posts likely differ in content to public ones, and their inclusion in the study would likely affect the descriptive results reported here.

In concluding the study, an important note must be made regarding the digital architectures framework: *digital architectures are subject to rapid and transformative change*. Even though Snapchat's architecture, for example, offered only rudimentary analytics to campaigns during the primaries, the platform was updated by the general election to provide campaigns with a sophisticated means of acquiring users' emails. The Trump campaign, says Oczkowski, gathered "hundreds of thousands of emails off the Snapchat platform" by presenting users with advertisements encouraging them to "swipe up" and enter their email addresses. Even in the interim between the 2016 primaries and the writing of this article, all of the platforms included here have undergone significant transformations in their digital architectures. Nevertheless, the comparison's purpose has been to elucidate how the architectures of a platform can be compared, systematically, at a particular point in time.

Future scholars may wish to engage with the question of how changes in a platform's digital architecture over time influence campaigning practices longitudinally, as well as how the architectures of platforms not analyzed here (e.g., YouTube or WhatsApp) affect campaigns' digital communication strategies. Moreover, data from other sources such as voter turnout, donation, or polling figures should be incorporated into future research designs, in order to corroborate how digital communication is impacted by offline dynamics critical for campaigns and their strategies.